\documentclass[twocolumn,preprintnumbers,amsmath,amssymb]{revtex4}  
\usepackage{graphicx}% Include figure files
\usepackage{dcolumn}% Align table columns on decimal point
\usepackage{bm}% bold math
\usepackage{multirow}
\usepackage{color}
\usepackage{amsmath}
\begin{document}
\title{Spin-orbit engineering in transition metal dichalcogenide alloy monolayers}

\author{Gang Wang$^1$}
\author{Cedric Robert$^1$}
\author{Aslihan Suslu$^2$}
\author{Bin Chen$^2$}
\author{Sijie Yang$^2$}
\author{Sarah Alamdari$^2$}
\author{Iann C. Gerber$^1$}
\author{Thierry Amand$^1$}
\author{Xavier Marie$^1$}
\author{Sefaattin Tongay$^2$}
%\email{Sefaattin.Tongay@asu.edu}
\author{Bernhard Urbaszek$^1$}
%\email{urbaszek@insa-toulouse.fr}

\affiliation{%
$^1$ Universit\'e de Toulouse, INSA-CNRS-UPS, LPCNO, 135 Av. Rangueil, 31077 Toulouse, France}
\affiliation{%
$^2$ School for Engineering of Matter, Transport and Energy, Arizona State University, Tempe, AZ 85287, USA}

\maketitle
%Nature intro paragraph
\textbf{Transition metal dichalcogenide (TMDC) monolayers are newly discovered semiconductors for a wide range of applications in electronics and optoelectronics \cite{Lopez:2013a,Xu:2014a,Zhang:2014a,Mak:2010a}. Most studies have focused on binary monolayers that share common properties \cite{Kormanyos:2015a}: direct optical bandgap, spin-orbit (SO) splittings of hundreds of meV, light-matter interaction dominated by robust excitons \cite{Ye:2014a,Ugeda:2014a,Wang:2015b} and coupled spin-valley states of electrons \cite{Mak:2014a,Xiao:2012a,Cao:2012a,Sallen:2012a}. 
Studies on alloy-based monolayers are more recent \cite{Tongay:2014a,Chen:2013a,Zheng:2015a,Kutana:2014a,Gan:2014a,Xi:2014a,Wei:2014a,Dumcenco:2012a,Feng:2014a,Chen:2014a,Feng:2015a}, yet they may not only extend the possibilities for TMDC applications through specific engineering but also help understanding the differences between each binary material.
Here, we synthesized highly crystalline Mo$_{1-x}$W$_{x}$Se$_2$ to show engineering of the direct optical bandgap and the SO coupling in ternary alloy monolayers. As the tungsten (W) composition is increased, we find a non-linear increase of the optically generated valley polarization degree and a strong impact on the temperature evolution of the photoluminescence (PL) emission intensity, introducing Mo$_{1-x}$W$_{x}$Se$_2$ monolayers as versatile building blocks for SO-engineered Van der Waals heterostructures \cite{Geim:2013a}.}\\
\begin{figure*}[ht!]
\includegraphics[width=0.75\textwidth]{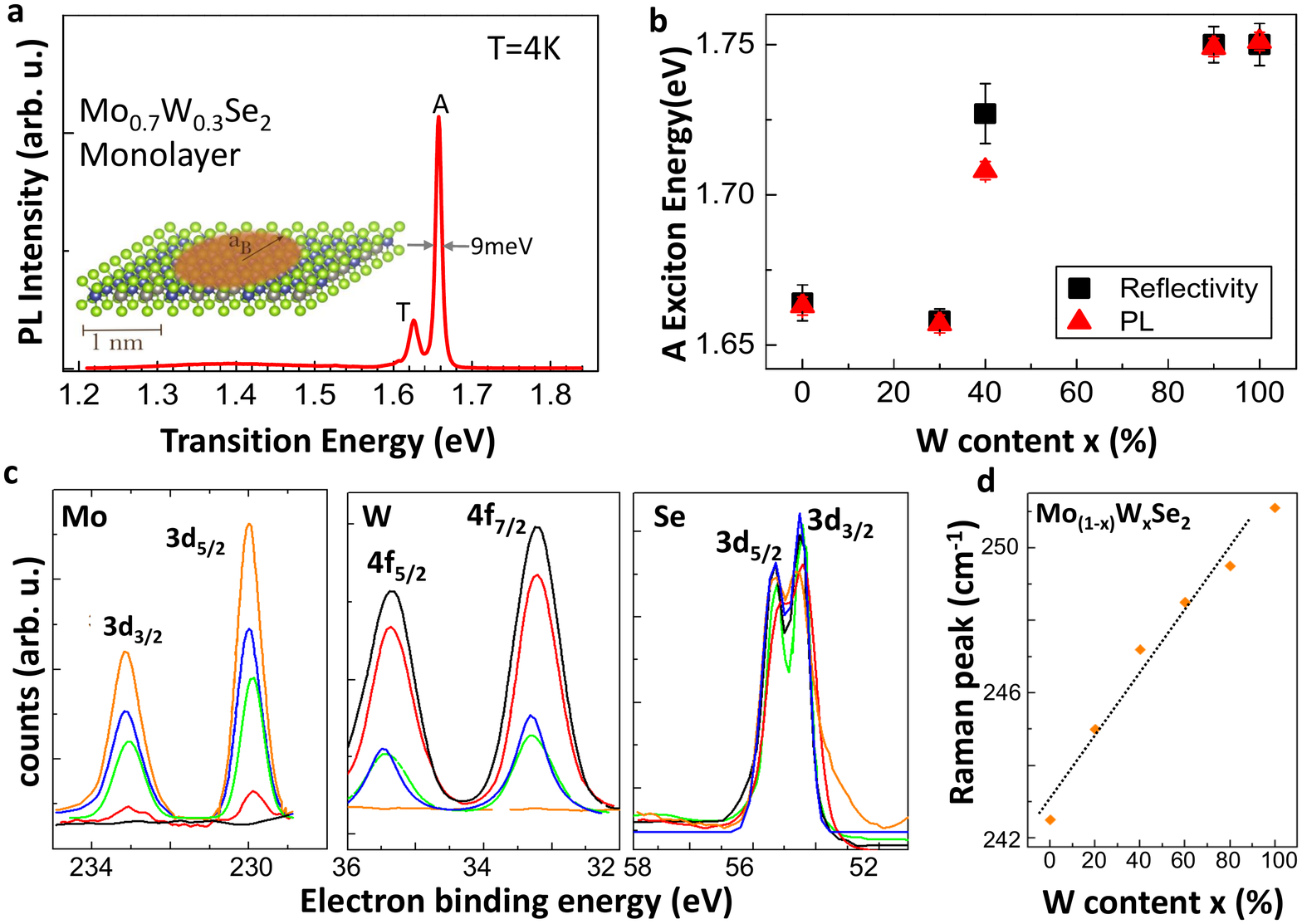}
\caption{\label{fig:fig1} \textbf{Highly crystalline Mo$_{1-x}$W$_{x}$Se$_2$ alloy monolayers} \textbf{(a)}  Low temperature photoluminescence (PL) spectroscopy is a very simple and efficient technique to probe the material quality. Impurities and defects will trap optically excited carriers, resulting in emission below the optical bandgap. PL spectrum at T=4~K of Mo$_{0.7}$W$_{0.3}$Se$_2$ alloy monolayer showing very sharp emission of the charged exciton (trion T) and the neutral A-exciton (A) and negligible defect related emission.  Inset: Representation of the alloy monolayer, the order of magnitude of the Bohr radius $a_B$ of an electron-hole pair (exciton) is shown. The narrow PL linewidth therefore indicates high quality alloy material on a nano-scopic scale. \textbf{(b)} A-exciton energy at T=4~K from PL (red triangles) and reflectivity measurements (black squares).\textbf{(c)} nano-resolution x-ray photoelectron (nano-XPS) measurements on Mo$_{1-x}$W$_{x}$Se$_2$ alloys showing gradual composition change with varying x, where orange, green, blue, red, black correspond to $x=0\%, 30\%, 40\%, 90\%$ and $100\%$, respectively.  For increasing x (W) content, W (Mo) content increases (decreases), whereas selenium ratio remains at the same values without any significant single or double (V$_{Se}$ and V$_{2Se}$) vacancy formation.  \textbf{(d)} $E_{2g}$ Raman peak position shift as a function of composition x.}
\end{figure*}
\indent The current generation of commercial optoelectronic devices based on III-V semiconductors owns its success to alloy engineering, adapting the optical and electronic properties of epitaxial layers by tuning the band structure and band alignment of each layer \cite{Rosencher:2002a}. Important examples include tuning the wavelength of light emitting devices and the absorption threshold in photovoltaic junctions. Tuning the SO coupling in the bands has an impact on reducing Auger effects in semiconductor lasers \cite{Adams:1986a} and on enhancing the spin-Hall effect for spintronic technologies \cite{Okamoto:2014a}.
Strong SO coupling in TMDCs causes a very large valence band spin splitting of several hundred meV \cite{Zhang:2014a,Riley:2014a}. This determines the energy position of the dominant A- and B-exciton resonances, for which optical absorption is of the order of 10\% \cite{Mak:2010a}. 
Because the SO coupling depends mainly on the transition metal d-orbitals, the splitting for WSe$_2$ ($\sim400$~meV \cite{Riley:2014a}) is roughly twice that of MoSe$_2$ ($\sim200$~meV \cite{Zhang:2014a}). This should allow for a wide tuning range in ternary alloys, which we demonstrate experimentally. The spin splitting in the conduction band is much smaller, but the sign depends on the transition metal \cite{Kormanyos:2015a}. This determines the nature of the A-exciton ground state as optically bright or dark (Fig.~\ref{fig:fig2}a) and has also important consequences for the spin and valley control.
By studying monolayers of high quality (Mo,W)Se$_2$ ternary alloys at various compositions we show non-linear evolutions for the bandgap, the SO-coupling and the optically generated valley polarization. \\
\begin{figure*}[ht!]
\includegraphics[width=0.99\textwidth]{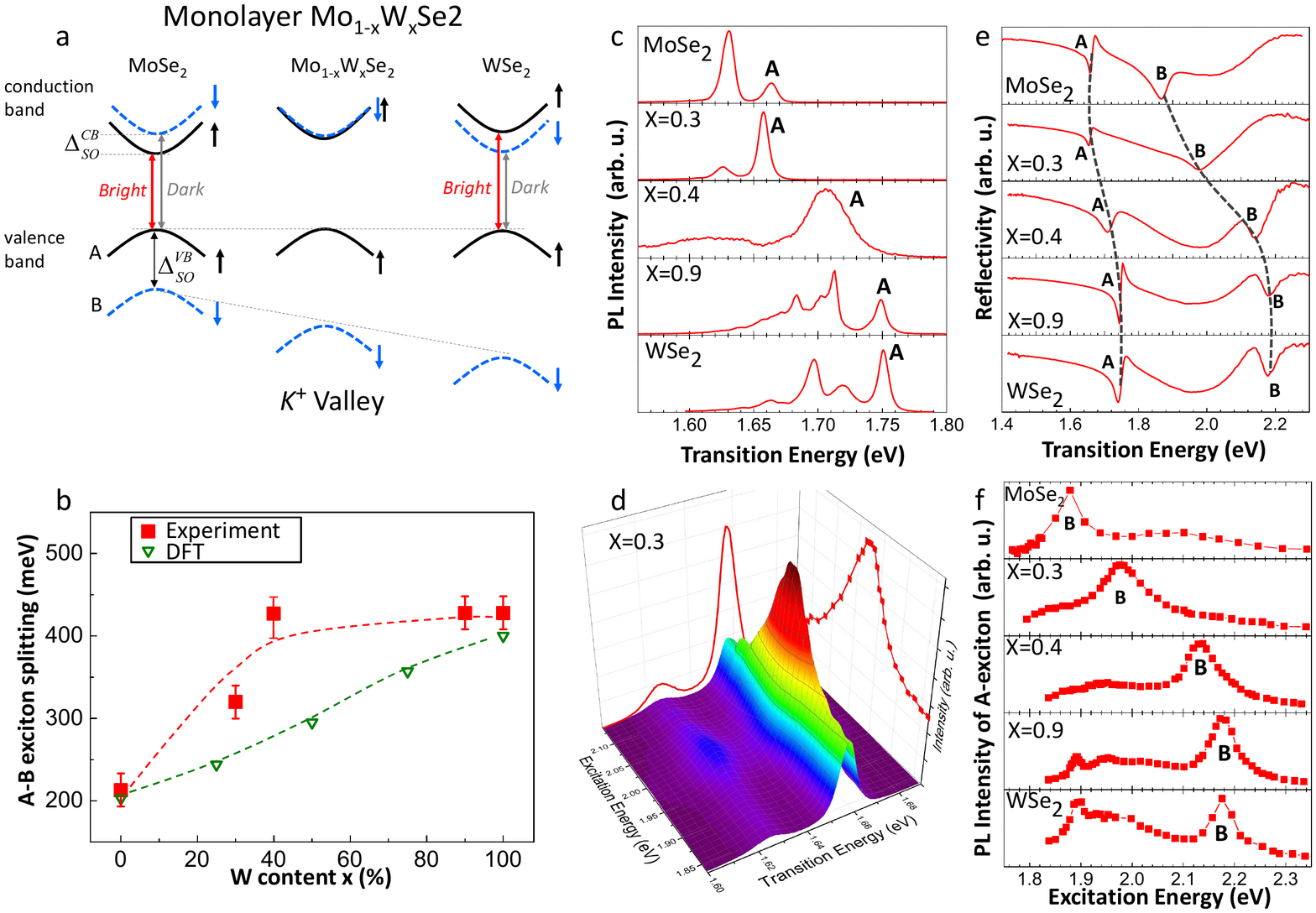}
\caption{\label{fig:fig2}  \textbf{Tuning the Spin-Orbit splitting in Mo$_{1-x}$W$_{x}$Se$_2$ monolayers} \textbf{(a)} Simple bandstructure scheme in the $K^+$ valley (at the $K$ point of the Brillouin zone) to indicate the different signs and magnitudes of the valence ($\Delta_{SO}^{VB}$) and conduction band spin splittings ($\Delta_{SO}^{CB}$) when going from MoSe$_2$ to WSe$_2$ monolayers. Optically bright (red arrows) and dark (grey arrows) A-exciton transitions are indicated \textbf{(b)} The splitting between A- and B- excitons, mainly given by the valence band SO-splitting, is measured by PLE (red squares), the dotted line is a guide to the eye. The sum of the valence band and conduction band spin splittings calculated by DFT is shown for comparison, for individual values and computational details see Methods and supplement. \textbf{(c)} PL spectra at 4K of monolayers for tungsten (W) composition from $x=0$ to 100 $\%$ in Mo$_{1-x}$W$_{x}$Se$_2$. The dominant, sharp A-exciton emission is indicated. \textbf{(d)} PL spectra of A-exciton in monolayer Mo$_{0.7}$W$_{0.3}$Se$_2$ for different laser excitation energies (PL excitation spectroscopy – PLE). We uncover a clear maximum when the laser energy is in resonance with the B-exciton. \textbf{(e)} Reflectivity spectra using a white light source, uncovering in addition to the A-exciton also the B-exciton spectral position, that can be tuned by varying the alloy composition. \textbf{(f)} Same measurements as d but for all samples, the A-exciton PL intensity is plotted as a function of laser energy. These PLE measurements allow determining the B-exciton energy with very high precision. 
}
\end{figure*} 
\indent The first challenge during crystal growth is to avoid phase separation and defect / cluster formation. This is a particular concern for materials containing anion atoms with large electronegativity or lattice constant differences. Owing to the rather close electronegativity and lattice constant values, as well as high miscibility, TMDCs alloys are stable at room temperatures \cite{Kang:2013a}. Previously, TMDCs alloys were synthesized either by mixing different chalcogen atoms (MX$_{2(1-n)}$Y$_{2n}$) or transition metals (M$_{(1-x)}$Z$_x$X$_2$) for broadly tunable optical band gaps using chemical vapour deposition (CVD) or conventional low pressure vapour transport (LPVT) techniques to yield materials in monolayer or bulk (layered) form \cite{Feng:2014a,Chen:2014a,Feng:2015a}.

In this work, Mo$_{1-x}$W$_{x}$Se$_2$ monolayers were exfoliated from their bulk counterparts which were grown using LPVT (see methods). 
We confirm the intended stoichiometry values using nano- x-ray photoelectron spectroscopy (nano-XPS), see Fig.~\ref{fig:fig1}c. 
Nano-XPS data taken from different spots closely match each other with 2\% W/Mo deviation and implies that synthesized materials are uniform in composition. For different $x$ values in Mo$_{1-x}$W$_{x}$Se$_2$ monolayers the most prominent in-plane Raman peak (E$_{2g}$) gradually shifts from 241 to 251~cm$^{-1}$ with slight deviation from linearity which can be explained by the modified random element iso-displacement (MREI) model \cite{Chen:2014a,Tongay:2014a,Sahin:2013a}.

Photoluminescence (PL) results shown in Fig.~\ref{fig:fig1}a confirm the high quality of the ternary monolayers. Remarkably we detect virtually no defect related emission in monolayer Mo$_{0.7}$W$_{0.3}$Se$_2$. The PL is dominated by two sharp peaks, the neutral exciton $A$ and the trion $T$, just as in the case of binary MoSe$_2$ \cite{Ross:2013a,Wang:2015a}. The A-exciton PL FWHM of only 9 meV is extremely narrow for a ternary alloy compared, for instance, with the typical PL FWHM of binary MoS$_2$ of about 50 meV \cite{Mak:2010a}. The narrow PL linewidth allows us to determine the energy position of the A-exciton as a function of tungsten content. In Fig.~\ref{fig:fig1}b we demonstrate band gap bowing as we deviate from a strictly linear extrapolation between the binary bandgaps, similar to the findings at room temperature \cite{Tongay:2014a} and predictions by theory \cite{Kutana:2014a,Gan:2014a,Xi:2014a}.

Reflectivity gives access to both A- and B-exciton energies, whose separation is determined mainly by SO coupling. Results are shown in Fig.~\ref{fig:fig2}e. Spectrally narrow transitions confirm the position of the A-exciton at the same energy as the PL measurements for $x=0.3$ and $x=0.9$, indicating that exciton localization (Stokes shift) on defects, for example, is negligible. The transition several hundred meV above the A-exciton is due to the B-exciton that clearly shifts as the tungsten content is increased. The sharp exciton resonances that dominate the reflectivity spectra  are an additional indication for high material quality. We also perform PL excitation spectroscopy (PLE) to access further details: When the laser is in resonance with the B-exciton, we observe a strong enhancement of the A-exciton PL emission intensity, as can be clearly seen for monolayer  Mo$_{0.7}$W$_{0.3}$Se$_2$ in Fig.~\ref{fig:fig2}d for a laser energy of 1.98~eV. We have performed experiments identical to Fig.~\ref{fig:fig2}d for all investigated samples. The results are presented in Fig.~\ref{fig:fig2}f using the A-exciton as the detection energy. We clearly see a shift in energy of the maxima associated to the B-exciton as the tungsten content is varied, in agreement with our reflectivity results.

\begin{figure*}[ht!]
\includegraphics[width=0.98\textwidth]{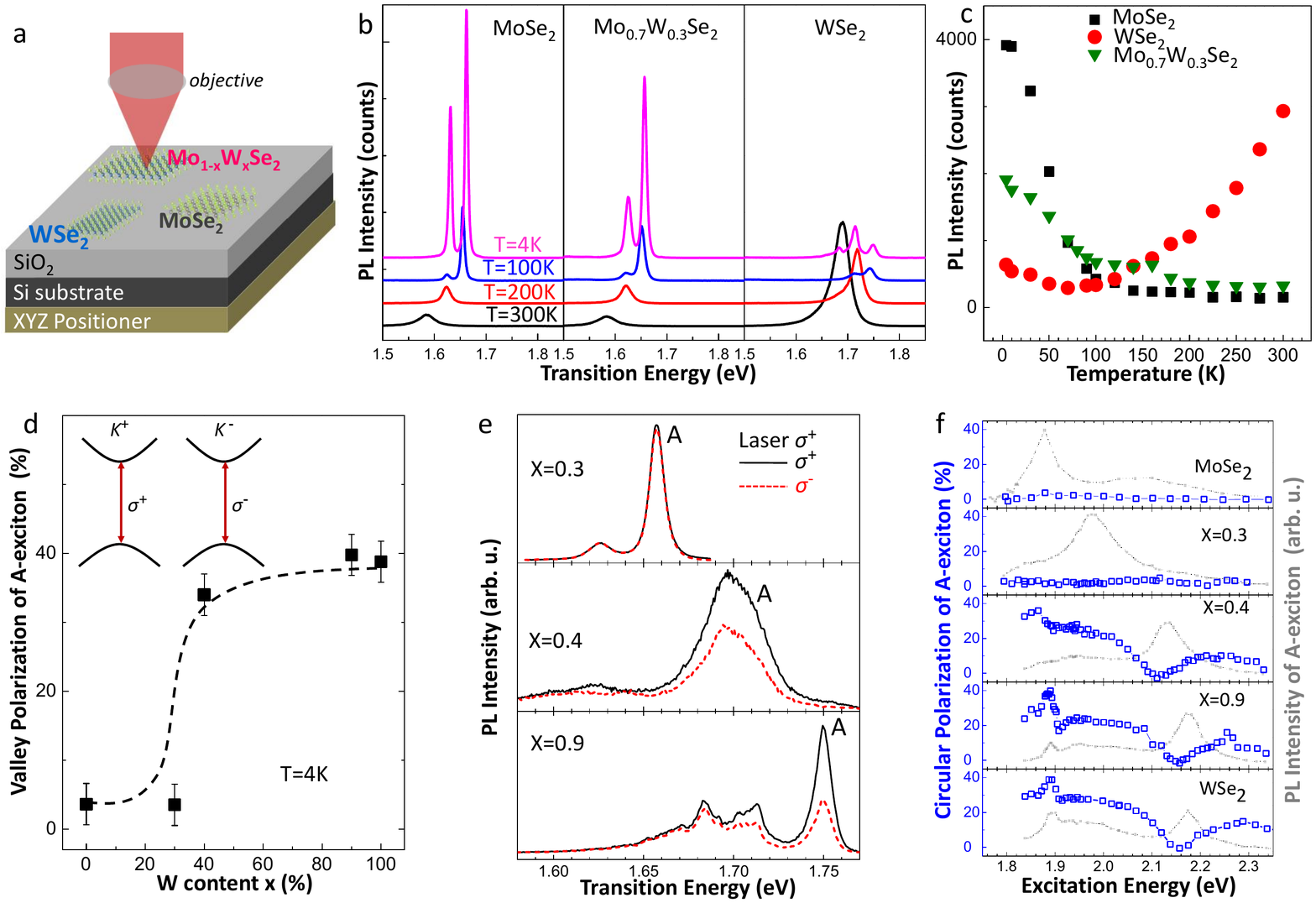}
\caption{\label{fig:fig3}  \textbf{Temperature dependence and Valley polarization engineering in Mo$_{1-x}$W$_{x}$Se$_2$ monolayers} \textbf{(a)} Comparing the global PL emission intensity of several samples is challenging, as the set-up has to compensate thermal expansion/movement. For the measurements of the PL emission intensity as a function of temperature we have exfoliated several monolayer flakes of different materials in close proximity onto the same substrate, which is mounted on a 3-axis attocube nano-positioner. Therefore PL emission for \textit{different} samples are measured under \textit{identical} conditions i.e. same detection and laser spot size.  Only highly reproducible in-plane movement is needed to change sample. \textbf{(b)} The PL spectra of monolayer  MoSe$_2$, Mo$_{0.3}$W$_{0.7}$Se$_2$ and WSe$_2$ are shown for different temperatures, the relative intensities can be directly compared.  \textbf{(c)}  Using the spectra from panel \textbf{b}, we integrate the total number of counts including A-exciton and trion (MoSe$_2$ and Mo$_{0.3}$W$_{0.7}$Se$_2$) and in addition the localized states (WSe$_2$). We compare the total number of counts for the three monolayer materials as a function of temperature, see supplement for details \textbf{(d)} The measured valley polarization i.e. circular PL polarization degree $P_c$ is plotted as a function of tungsten (W) content in the sample. $P_c$ is defined as $P_c=\frac{I^+-I^-}{I^++I^-}$, where $I^+$ and $I^-$ are the $\sigma^+$ and $\sigma^-$ polarized PL components, respectively. We observe a highly non-linear increase in the valley polarization as more tungsten is incorporated. For the measurement, for each sample the laser energy is 140~meV above the A-exciton \textbf{(e)} Using $\sigma^+$ circularly polarized laser excitation, we detect the A-exciton emission in $\sigma^+$ (black) and $\sigma^-$ (red) polarization. For $x=0.3$ we detect no polarization, as for binary MoSe$_2$. Surprisingly, for $x=0.4$ we detect up to $40 \%$ PL polarization. The results for $x=0.9$ also show high polarization. \textbf{(f)} The circular PL polarization $P_c$ is plotted as a function of the excitation laser energy to find optimal valley polarization conditions. Whereas for $x\leq0.3$ the valley polarization remains low, we demonstrate for $x\geq0.4$ a wide range of laser excitation energies that can be used for valley index initialization.
}
\end{figure*}

The values of the A-B exciton splitting measured with two independent techniques are summarized in Fig.~\ref{fig:fig2}b.  For Mo$_{0.7}$W$_{0.3}$Se$_2$ monolayers we measure a separation of 320~meV, an increase of 50\% compared to binary MoSe$_2$.  For Mo$_{0.6}$W$_{0.4}$Se$_2$ the separation is about 420~meV, very close to the value of 425~meV we measure for binary WSe$_2$. Surprisingly, our experiments show a clear SO bowing, which has not been anticipated yet by theory as spin-orbit effects where not included in references \cite{Kutana:2014a,Gan:2014a,Xi:2014a}. Spin orbit bowing has been observed previously in III-V semiconductor alloys \cite{Wei:1989a, Fluegel:2006a}. Here our experimental input for TMDC alloys can be used in future models as a sensitive fingerprint of contributions from different orbitals to the SO coupling.
For order of magnitude estimations we have calculated valence and conduction band splittings (see supplement) by density functional theory DFT using an artificial supercell with just 4 transition metal atoms in ordered configurations. We show their sum for comparison with the experiments in Fig.~\ref{fig:fig2}b, the influence of the realistic local lattice symmetry \cite{Wei:1989a,Kutana:2014a,Gan:2014a,Xi:2014a}, possibly different A- and B-exciton binding energies and effective masses is not included \cite{Kormanyos:2015a}. The SO splitting is a key material parameter, also for applications demanding strong broadband absorption in the visible such as photodetectors and solar cells. We demonstrate that by employing different ternary TMDC alloys, not only the absolute energy but also the relative separation of  the A- and B-excitons can be tuned over hundreds of meV.

The sign and amplitude of the SO-splitting in the conduction band will strongly influence the balance between optically dark and bright transitions in WSe$_2$, MoSe$_2$ and ternary alloys (Fig.~\ref{fig:fig2}a). This has a direct impact on the light emission yield from cryogenic to room temperatures, the SO coupling therefore needs to be controlled for optoelectronics applications. Our experiments allow us to uncover important differences between WSe$_2$, MoSe$_2$ and the ternary samples, as can be seen directly when comparing the PL intensities in Fig.~\ref{fig:fig3}b. 
For monolayer WSe$_2$ we detect comparatively weak emission at 4~K, increasing by one order of magnitude when going to T=300~K (Fig.~\ref{fig:fig3}c), a trend also observed in \cite{Arora:2015a}. 
One possible explanation for the increase of the PL intensity as a function temperature is the thermal conversion of dark into bright states, as observed for CdSe nano-crystals \cite{Crooker:2003a}, where dark states also lie energetically below bright states. In stark contrast, for binary MoSe$_2$ monolayers we find a drastic decrease of the PL emission intensity when going from T=4~K to 150~K, consistent with the opposite order of bright and dark states compared to WSe$_2$, see Fig.~\ref{fig:fig2}a. The ternary sample Mo$_{0.3}$W$_{0.7}$Se$_2$ shows a behaviour qualitatively similar to binary MoSe$_2$, albeit with much stronger emission for temperatures from 100 to 200~K. For temperatures between 4 and 100~K we are able to extract separately the PL intensity evolution of the trion and the neutral A-exciton for binary and ternary samples, see supplement for full details. For binary MoSe$_2$ we find that the trion intensity decreases much faster with temperature than the neutral exciton.
In contrast, for monolayer Mo$_{0.3}$W$_{0.7}$Se$_2$ the trion and A-exciton PL intensity show similar temperature dependence. For a fully quantitative analysis of the global PL yield evolution, non-radiative channels due to material imperfections need to be taken into account in addition to the dark-bright state competition discussed here. We have confirmed the stark contrast between MoSe$_2$ and WSe$_2$ for monolayers not only from our LPVT samples but also from commercial material, which shows the same behaviour (see supplement).

We also investigate the impact of tuning the sign and amplitude of the SO-coupling on spin- and valley physics in TMDC monolayers. The electron \textit{valley} degree of freedom is accessible in monolayer TMDCs in simple optical manipulation schemes \cite{Xiao:2012a,Cao:2012a,Sallen:2012a} and as is coupled to the electron spin degree of freedom. For this reason valleytronics is one of the main research directions in TMDCs \cite{Mak:2014a,Xu:2014a}. Energetically degenerate states in the $K^+$ and $K^-$ valleys have opposite spins due to time reversal symmetry \cite{Xiao:2012a,Cao:2012a}. Large SO splittings will also stabilize the valley index, as spin flips are necessary to change valley in momentum space. Although MoSe$_2$ and WSe$_2$ show comparable structural and optical quality \cite{Ross:2013a,Jones:2013a,Macneill:2015a}, their valley polarization properties are very different: In WSe$_2$ not only valley polarization can be generated, but also a superposition of $K^+$ and $K^-$ states, termed valley coherence \cite{Jones:2013a}. In contrast, the valley polarization of MoSe$_2$ at zero magnetic field rarely exceeds 5\% using various optical initialization schemes \cite{Macneill:2015a,Wang:2015a}.  Here we investigate how valley polarization generation improves in TMDC alloys as we go from MoSe$_2$ to WSe$_2$. For the high quality $x=0.3$ sample, we do not detect measurable valley polarization when scanning a very broad range of laser energies, as can be seen in Fig.~\ref{fig:fig3}e and Fig.~\ref{fig:fig3}f. By increasing the tungsten content by only 10\% to $x=0.4$ we can increase the valley polarization by an order of magnitude to about 40\%. 
An efficient way for successful optical valley initialization of the A-exciton is to tune the excitation laser into resonance with exciton ground or excited states \cite{Wang:2015b}. For MoSe$_2$ the B-exciton 1s and the A-exciton 2s/2p states overlap energetically \cite{Wang:2015c}, which might be one of the reasons that prevents optical valley initialization. For WSe$_2$ the situation is much simpler, as the B-excion is at much higher energy compared to the excited A-exciton states. For $x=0.4$ we are again in this favourable situation, which might be beneficial for optical valley initialization.

In summary, we show clear tuning of the energy and separation of the dominant exciton transitions  A and B at the direct bandgap in high quality Mo$_{1-x}$W$_{x}$Se$_2$ alloy monolayers. We investigate the impact of the tuning of the SO spin splitting on the optical and polarization properties. We show a non-linear increase of the optically generated valley polarization as a function of tungsten concentration, where 40\% tungsten incorporation is sufficient to achieve valley polarization as high as in binary WSe$_2$. We also probe the impact of the tuning of the conduction band SO spin splitting on the bright versus dark state population i.e. PL emission intensity. We show that the MoSe$_2$ PL intensity decreases as a function of temperature by an order of magnitude, whereas for WSe$_2$ we measure surprisingly an order of magnitude increase over the same temperature range ($T=4-300$~K).  The ternary material shows a trend between these two extreme behaviours. These results show the strong potential of SO engineering in ternary TMDC alloys for optoelectronics and applications based on electron spin- and valley-control.

\textbf{METHODS.}  
\textit{Sample growth and characterization.---}
Layered Mo$_{1-x}$W$_{x}$Se$_2$ semiconductors were grown by modified low-pressure vapour transport (LPVT) technique to achieve high optical quality materials. Synthesized materials are fully alloyed (not phase separated) as confirmed by three complementary techniques, such as sub-$\mu$ Raman spectroscopy, sub-micron photoluminescence, and nano-XPS (x-ray photoelectron spectroscopy), see Fig.~\ref{fig:fig1}. X-ray diffraction measurements  display sharp ($\sim 1~\mu$m domain size) (001) peaks. Thus TMDCs alloys are highly layered and crystallized in the hexagonal phase, and contains  negligible amount of minority crystal orientation. Observed (001) peaks of Mo$_{1-x}$W$_{x}$Se$_2$ appear at the same position (within $<0.08^\circ$ ) implying that alloys have similar c-axis parameters. Consistent with nano-XPS measurements, Rutherford backscattering spectroscopy (RBS) data collected from $\sim 100~\mu$m  diameter circle agrees well with the XPS measurements (not shown).  \\
\textit{Optical Spectroscopy.---}
Experiments at $T=4-300$~K are carried out in a confocal microscope \cite{Wang:2015b}. The detection spot diameter is $\approx1\mu$m, i.e. considerably smaller than the monolayer flake size of $\sim 10~\mu$m$\times10~\mu$m. For time integrated experiments, the PL emission is dispersed in a spectrometer and detected with a Si-CCD camera. The samples are excited either by a cw He-Ne laser or ps pulses generated by a tunable frequency-doubled optical parametric oscillator (OPO) synchronously pumped by a mode-locked Ti:Sa laser. The typical pulse temporal and spectral widths are 1.6 ps and 3 meV, respectively, the repetition rate is 80 MHz.  \\
\textit{Theory.---} The electronic structures of Mo$_{1-x}$W$_{x}$Se$_2$ with $x=0.25, 0.50, 0.75$ were simulated within the Density Functional Theory (DFT) framework as implemented in the VASP\cite{Kresse:1996a} package. Order of magnitude trends of the conduction and valence spin-splitting of these elemental compositions were simply computed in a (2$\times$2) supercell, containing 4 transition metal atoms in total, and neglecting for simplicity random and more realistic configurations,see supplement.

\textbf{Acknowledgements} We acknowledge funding from ERC Grant No. 306719, ANR MoS2ValleyControl. I. C. G. acknowledges the CNRS for financial support and CALMIP initiative for the generous allocation of computational times, through the project p0812, as well as the GENCI-CINES and GENCI-IDRIS for the grant x2015096649. S.T acknowledges Arizona State University start-up grant and Ira. A. Fulton College of Engineering Origami Seeding Funds.\\ 

\section{Supplement}
\subsection{Optical Spectroscopy}
We analyse separately the evolution with temperature of the neutral A-exction and trion PL emission intensity by careful fitting multi-Gaussian lineshapes for ML MoSe$_2$,  Mo$_{0.7}$W$_{0.3}$Se$_2$ and WSe$_2$. Here we concentrate on the temperature range where trions and excitons can be clearly distinguished. The bold, large symbols correspond to experiments on our LPVT grown samples, the smaller, semitransparent symbols show for comparison measurements on commercial (2D Semiconductors) binary material. Note that the A-exciton to trion intensity ratio for each sample will depend on the excess carrier concentration. In our samples the residual doping level cannot be controlled, for more deterministic charge control gated samples are needed \cite{Ross:2013a}.
\begin{figure}[t]
\includegraphics[width=0.45\textwidth]{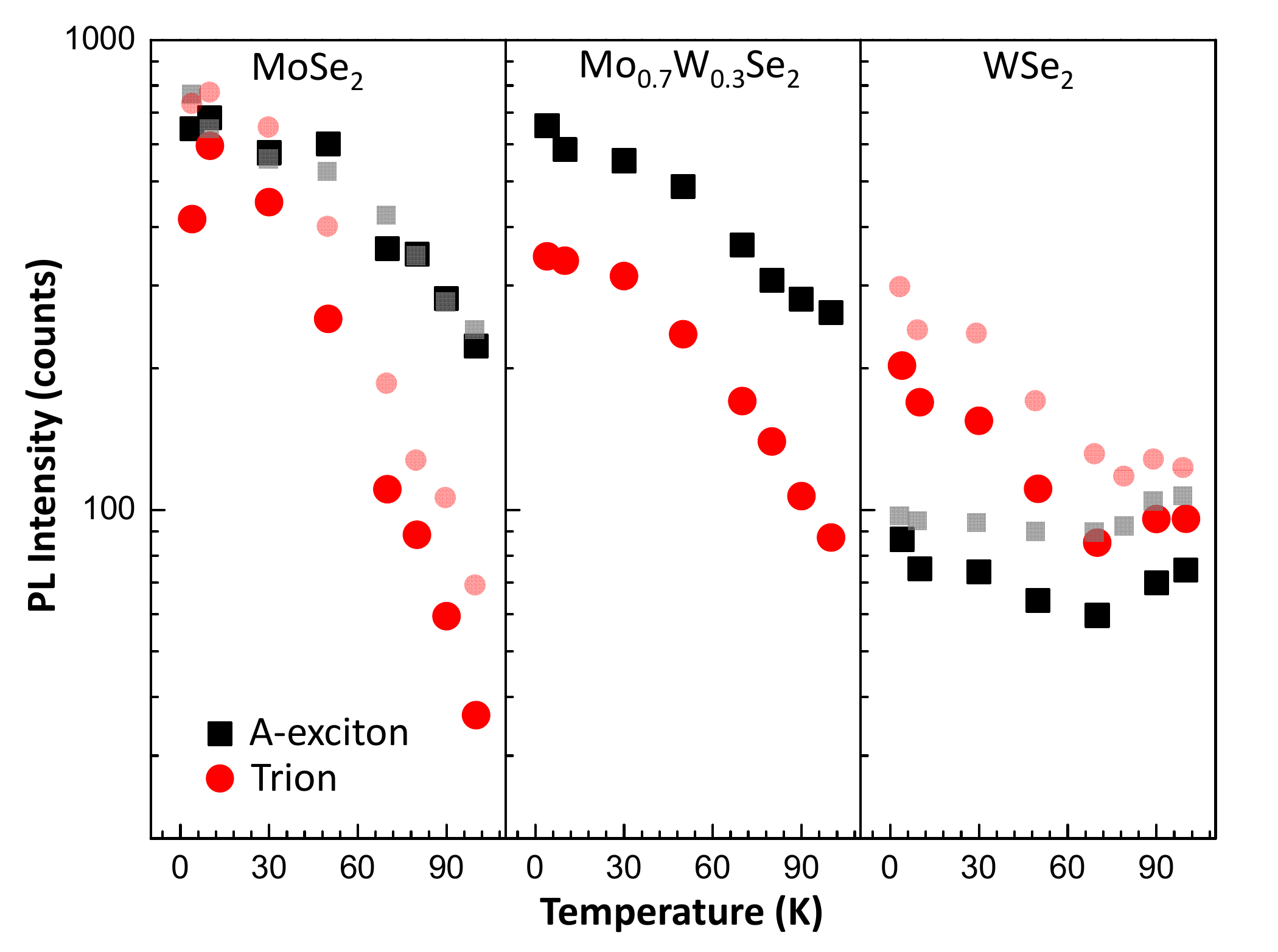}
\caption{\label{fig:figSI0} Evolution with temperature of the neutral A-exction (black squares) and trion(red circles) PL emission intensity for ML MoSe$_2$,  Mo$_{0.7}$W$_{0.3}$Se$_2$ and WSe$_2$. Bold, large symbols correspond to experiments on our LPVT grown samples, the smaller, semitransparent symbols show for comparison measurements on commercial (2D Semiconductors) binary material.  }
\end{figure}
To go further in our direct comparison of PL emission intensities for different samples, we plot in Fig.~\ref{fig:figSI1} the ratio of the PL intensity of WSe$_2$ divided by the intensity of MoSe$_2$ measured during the same experiment. We observe for our LPVT grown samples and commercial samples the same trend. The MoSe$_2$ emission gets weaker as temperature increases, whereas WSe$_2$ emission intensity gets stronger. Very similar results for samples grown under different conditions indicate that the PL evolution is due to intrinsic effects such as the competition between spin-orbit split bright and dark states \cite{Dery:2015a}, and cannot merely be attributed to different sample quality. For the ternary sample we see a clear difference to the MoSe$_2$ evolution between $T=100$~K and 200~K.\\
\begin{figure}[t]
\includegraphics[width=0.45\textwidth]{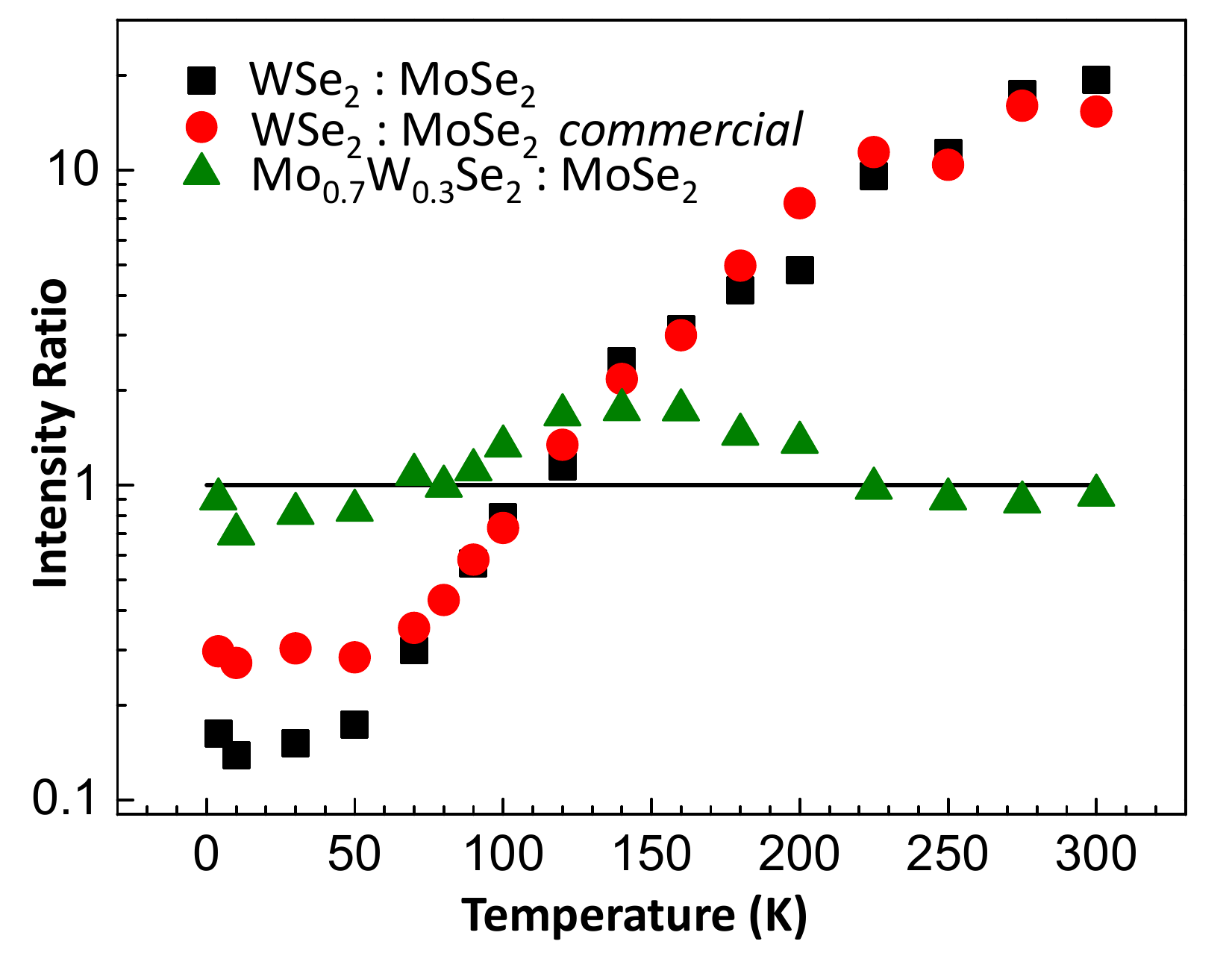}
\caption{\label{fig:figSI1} PL emission intensity of three W containing samples is divided by the MoSe$_2$ PL intensity. The interest in plotting the ratio at each temperature is to eliminate any difference in detection efficiency that exist for different set-up temperatures. Commercial WSe$_2$ samples (red circles) show the same trend as our LPVT grown samples (black squares). Alloy Mo$_{0.3}$W$_{0.7}$Se$_2$ : MoSe$_2$ PL ratio shown as green triangles.}
\end{figure}
\begin{figure*}[ht!]
\begin{center}
a)
\begin{minipage}[c]{0.4\textwidth}
\includegraphics[width=0.9\textwidth]{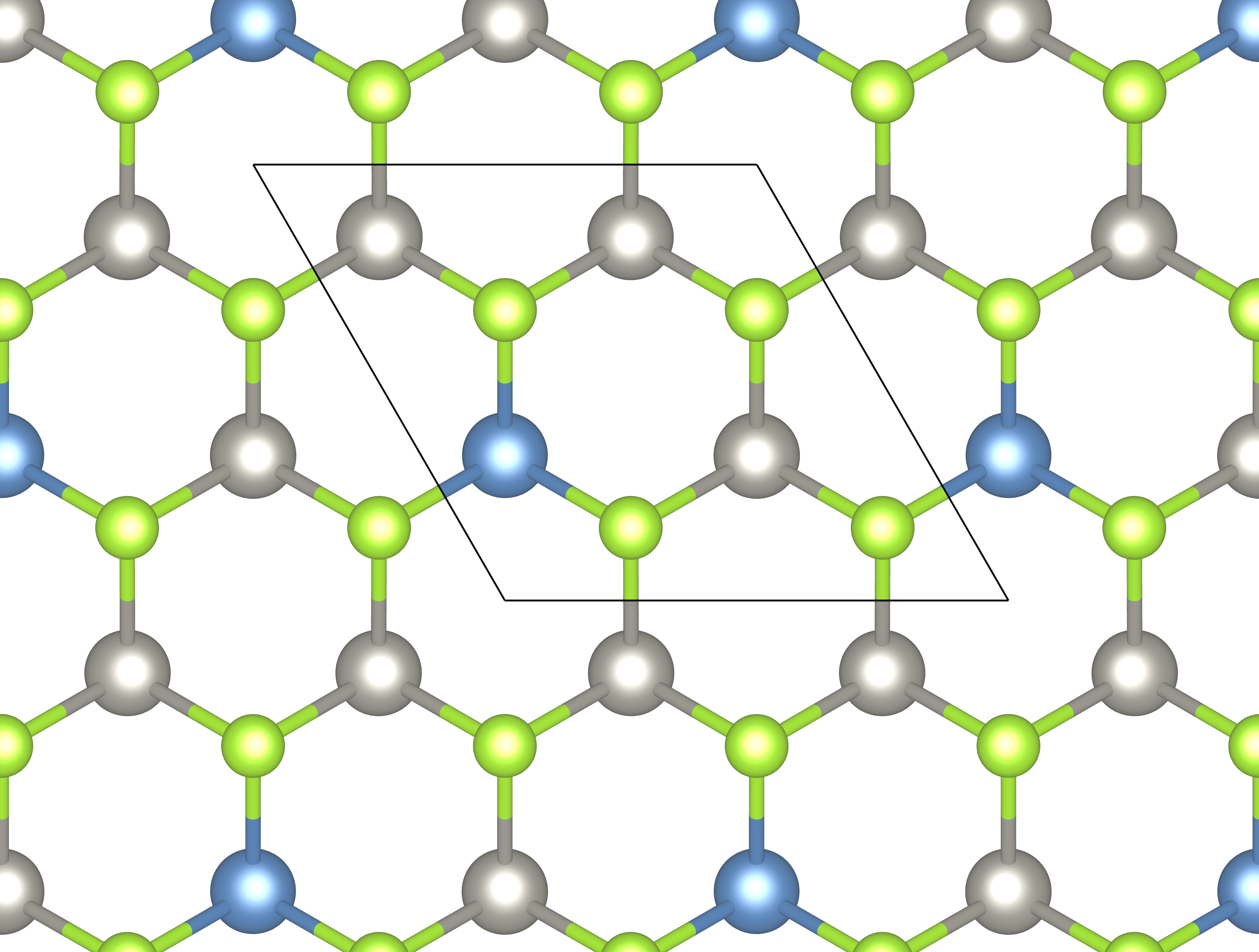}
\end{minipage}
b)
\begin{minipage}[c]{0.4\textwidth}
 \includegraphics[width=0.9\textwidth]{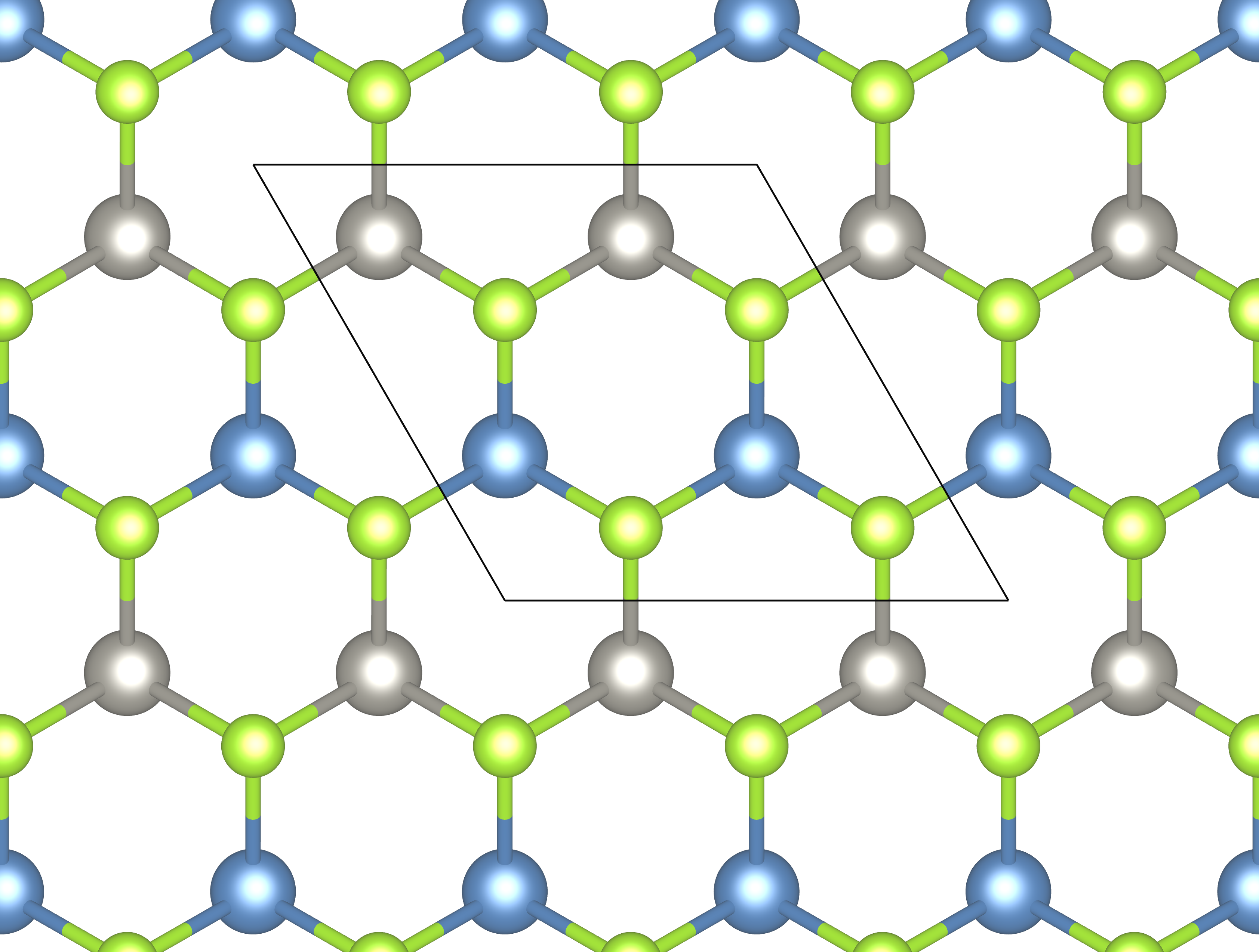}
 \end{minipage}
\end{center}
\caption{\label{fig:figSI2} Ordered configurations used for the calculations with (a) x=0.25 and (b) x=0.5. The calculation cell is given with the black line. Mo atoms are in blue, W atoms in grey and Se in green}.
\end{figure*}
\subsection{Density Functional Theory}
\subsubsection*{Computational Details}
First-principles calculations are performed using the Vienna Ab-initio Simulation Package~\cite{Kresse:1993a,Kresse:1994a,Kresse:1996a,Kresse:1996b} and the Perdew-Burke-Ernzerhof generalized gradient approximation~\cite{Perdew:1996a} for exchange-correlation functional. It uses the plane-augmented wave scheme \cite{Blochl:1994a, Kresse:1999a} to treat core electrons. Fourteen and six electrons for metals and Se respectively are explicitly included in the valence, with an energy cut-off of 400 eV. All atoms are allowed to relax with a force convergence criterion below $0.005$ eV/\AA, using experimental lattice parameters of 3.289 and 3.282 \AA~for the MoSe$_2$-based and for WSe$_2$-based cell respectively. Spin-Orbit Coupling (SOC) is included non-self consistently, since for these calculations the valence band splitting deviates by less than 1 meV from the values obtained from self-consistent SOC calculations on primitive cells.  A grid of 12$\times$12$\times$1 and 6$\times$6$\times$1 k-points have been used for primitive and (2$\times$2) cell respectively, in conjunction with a vacuum height of 17 \AA. A gaussian smearing with a width of 0.05 eV is used for partial occupancies, when a tight electronic minimization tolerance of $10^{-8}$ eV.\\
\subsubsection*{Configurations}
Random configurations in the lattice of Mo$_{1-x}$W$_{x}$Se$_2$ alloys will have an impact on the experiments in the present study and on previous ones for Mo$_x$W$_{(1-x)}$S$_2$~~\cite{Dumcenco:2012a, Chen:2013a} In our simple approach, the determination of the spin-splitting due to Spin-Orbit Coupling (SOC) in Conduction Band (CB) and Valence Band (VB) is only possible for ordered structures since supercell calculations suffer from band folding in the first Brillouin zone.
Energetic aspects have theoretically studied in details for disulphide compounds mainly, but Gan \textit{et al} have reported, based on DFT calculations, that Mo-Se-W bonds are slightly lower in energy than Mo-Se-Mo or W-Se-W bonds \cite{Gan:2014a}, a statement also confirmed by another work \cite{Wei:2014a}. Besides in ref.\cite{Xi:2014a}, it is shown that when the percentage of W is increased the hole effective mass decreases linearly and more importantly the electron effective mass of alloys is always larger than that of the pure systems. Since transition metal {\textit{d}}-orbitals contribute differently to conduction bands for MoS$_2$ and WS$_2$, but do not to valence bands, it is clear that electrons and holes in alloys behave differently when compared to pure systems. Considering band gaps, a slight bowing has been also reported at the LDA level, up to a W concentration of 0.5 \cite{Kutana:2014a}.  In the present study, we have considered only ordered configurations based on a (2x2) supercell, thus neglecting the randomization effects of the site occupations. Figure \ref{fig:figSI2} shows these ordered atomic configurations for $x=0.25$ and $x=0.5$ that have been used to extract spin splitting energy values in the valence and the conduction bands for two different values of lattice parameter.\\
\subsubsection*{Lattice parameter effects}
Influence of the lattice parameter choices on the spin-splitting of valence and conduction bands are presented in Table \ref{tab:tabSI1}.
Using a slightly larger lattice parameter, i.e the Mo-cell one, provides larger spin-splitting energy in the valence bands, but smaller ones in the conduction bands. All together, the separation of the A and B exciton energies is mainly governed by the spin-splitting, in addition different effective masses and exciton binding energies can contribute.
\begin{table}[h!]
\begin{center}
\begin{tabular}{lcccccccc}
\hline
\hline
& & \multicolumn{3}{c}{Exp. Mo-cell} & & \multicolumn{3}{c}{Exp. W-cell}     \\
& & \multicolumn{3}{c}{3.289 \AA} & & \multicolumn{3}{c}{3.282 \AA}     \\ 
 %& & & &  & & & &  & & \\
 %& & $\Delta_{SO}^{VB}$ & & $\Delta_{SO}^{CB}$ & \\
 \cline{3-5} \cline{7-9}\\
& &$\Delta_{SO}^{VB}$ & & $\Delta_{SO}^{CB}$ & &$\Delta_{SO}^{VB}$ & & $\Delta_{SO}^{CB}$ \\ 
x &\\
\hline
0.0 & & 452 & & +46 & & 449 & & +49 \\ 
\\
0.25 & & 375 & & +7 & & 372 & & +15 \\
\\
0.5 & & 305 & & +10 & & 304 & & +9 \\ 
\\
0.75  & & 242 & & -5 & & 241 & & -3 \\ 
\\
1.0 & & 182 & & -22 & & 181 & & -22 \\ 
%0.25 & & & & & & \\ 
%0.50 & & & & & & \\ 
%0.75 & & & & & & \\ 
%1.0 & & & & & & \\ 
\hline
\hline
\end{tabular}
\end{center}
\caption{\label{tab:tabSI1}Spin splitting energy values in the valence and the conduction bands in meV as function of the lattice parameter.}
\end{table}

\end{document}